\def\rn{\noindent\parshape 2 0.5truecm 8.5truecm 0.3truecm 8.2truecm}
\def\nn#1 #2{#2. #1}				
\def\nnn#1 #2 #3{#2. #3. #1}			
\def\nnnn#1 #2 #3 #4{#2. #3. #4 #1}		
\def\nnnnn#1 #2 #3 #4 #5{#2. #3. #4 #5. #1}	
\def\dualand{ and\hbox{ }}				
\def\multiand{, and\hbox{ }}				
\def\rf#1;#2;#3;#4;#5 {{\frenchspacing\par\rn#1, #3 {\bf #4}, #5 (#2). \par}}
\def\rg#1;#2;#3;#4;#5;#6 {{\frenchspacing\par\rn#1, #3 {\bf #4}, #5 (#2). \par}}
\def\rfbook#1;#2;#3;#4;#5 {{\frenchspacing\par\rn#1, {\it #3} (#5, #4, #2).\par}}
\def\rfprep#1;#2;#3 {{\par\frenchspacing\rn#1, #3 (#2).\par}}
\def\preskip {\vskip-0.0cm}
\def\postskip{\vskip+0.1cm}

\def\beq#1{\begin{equation}\label{#1}}
\def\eeq{\end{equation}}
\def\beqa#1{\begin{eqnarray}\label{#1}}
\def\eeqa{\end{eqnarray}}
\def\eq#1{equation~(\ref{#1})}

\def\draft{
}
\def\spose#1{\hbox to 0pt{#1\hss}}
\def\simlt{\mathrel{\spose{\lower 3pt\hbox{$\mathchar"218$}} \raise 2.0pt\hbox{$\mathchar"13C$}}}
\def\simgt{\mathrel{\spose{\lower 3pt\hbox{$\mathchar"218$}} \raise 2.0pt\hbox{$\mathchar"13E$}}}
\def\simpropto{\mathrel{\spose{\lower 3pt\hbox{$\mathchar"218$}} \raise 2.0pt\hbox{$\propto$}}}

\def\bt{\begin{tabbing}}
\def\et{\end{tabbing}}
\def\beq#1{\begin{equation}\label{#1}}
\def\eeq{\end{equation}}

\def\sec#1{Section~\ref{#1}}

\def\bfig{\begin{figure}[h] \centerline{\hbox{}}\vfill}
\def\efig{\end{figure}\vfill\newpage}

\def\fig#1{Figure~\ref{#1}}

\def\fig#1{Figure~\ref{#1}}
\def\Fig#1{Figure~\ref{#1}}

\def\lmax{{\ell_{\rm max}}}

\def\expec#1{\langle#1\rangle}
\def\tr{\hbox{tr}\>}
\def\l{\ell}
\def\etal{{\frenchspacing\it et al.}}
\def\ie  {{\frenchspacing\it i.e.}}

\def\etc {{\frenchspacing\it etc.}}

\def\A{{\bf A}}
\def\x{{\bf x}}
\def\r{{\bf r}}

\def\ns{n_s}
\def\nt{n_t}
\def\As{A_s}
\def\At{A_t}
\def\fn{f_\nu}
\def\Ok{\Omega_{\rm k}  }
\def\Ol{\Omega_\Lambda  }
\def\Oc{\Omega_{\rm cdm}}
\def\On{\Omega_\nu      }
\def\Ob{\Omega_{\rm b}  }
\def\Od{\Omega_{\rm dm} }
\def\ob{\omega_{\rm b}  }
\def\od{\omega_{\rm dm} }
\def\R{{\bf R}}

\def\bs{}

\def\A{{\bf A}}
\def\B{{\bf B}}
\def\C{{\bf C}}

\def\F{{\bf F}}

\def\I{{\bf I}}

\def\NN{{\bf\Sigma}}

\def\Q{{\bf Q}}

\def\R{{\bf R}}
\def\SS{{\bf S}}

\def\C{{\bf C}}
\def\Pr{{\bf \Pi}}
\def\P{{\bf P}}
\def\Q{{\bf Q}}
\def\boom{{\rm BOOMERanG}}
\def\q{{\bf q}}

\def\norm{{\cal N}}
\def\dT{\delta {\rm T}}

\def\dl{\Delta  \l}
\def\sigl{\sigma_{\l}}
 
\documentstyle[prd,aps,epsf]{revtex}
\begin{document}
\draft
\twocolumn[\hsize\textwidth\columnwidth\hsize\csname@twocolumnfalse\endcsname


\title{FIRST ATTEMPT AT MEASURING THE CMB CROSS-POLARIZATION}

\author{Ang\'elica de Oliveira-Costa$^{1}$, 
                         Max Tegmark$^{1}$, 
		  Matias Zaldarriaga$^{2}$,
		       Denis Barkats$^{3}$, 
		   Josh O. Gundersen$^{4}$, \\ 
		      Matt M. Hedman$^{5}$,
		   Suzanne T. Staggs$^{3}$  \& 
		      Bruce Winstein$^{5}$}
		      
\address{$^{1}$Department of Physics \& Astronomy, University of Pennsylvania, Philadelphia, PA 19104, USA, angelica@higgs.hep.upenn.edu\\		      
	 $^{2}$Department of Physics, New York University, New York, NY 10003, USA \\		      
	 $^{3}$Department of Physics, Princeton University, Princeton, NJ 08544, USA \\
	 $^{4}$Department of Physics, University of Miami, Coral Gables, FL 33146, USA \\
	 $^{5}$Center for Cosmological Physics, University of Chicago, Chicago, IL 60637, USA \\ 		      
	 }

 \date{\today. To be submitted to Phys. Rev. D.}

\maketitle

\begin{abstract}
We compute upper limits on CMB cross-polarization by cross-correlating the
PIQUE and Saskatoon experiments. We also discuss theoretical and practical 
issues relevant to measuring cross-polarization and illustrate them with 
simulations of the upcoming $\boom$ 2002 experiment. 
We present a method that separates all six polarization power spectra 
(TT, EE, BB, TE, TB, EB) without any other ``leakage'' than the familiar 
EE-BB mixing caused by incomplete sky coverage.  Since E and B get mixed, 
one might expect leakage between TE and TB, between EE and EB and between 
BB and EB  --- our method eliminates this by preserving the parity symmetry 
under which TB and EB are odd and the other four power spectra are even.
\end{abstract}

\pacs{98.62.Py, 98.65.Dx, 98.70.Vc, 98.80.Es}

\keywords{cosmic microwave background -- methods: data analysis}

] 


\section{INTRODUCTION}

Although not yet detected, it is theoretically expected that the 
Cosmic Microwave Background (CMB) is polarized. CMB polarization 
should be induced via Thomson scattering which occurs either at 
decoupling or during reionization. The level of 
this polarization is linked to the local quadrupole anisotropy 
of the incident radiation on the scattering electrons, and it is 
expected to be of order 1\%-10\% of the amplitude of the temperature 
anisotropies depending on the angular 
scale (see \cite{ZH95,HW97} and references therein).

CMB polarization is important for two reasons: first, 
polarization measurements can substantially improve the accuracy 
with which cosmological parameters are measured by breaking the degeneracy 
between certain parameter combinations; second, it offers 
an independent test of the basic assumptions that underly the 
standard cosmological model. 

Since the polarized CMB signal is so small, it is quite likely 
that its first detection will be an indirect statistical one, from
its predicted correlation with an unpolarized CMB map.
Since such cross-correlations involve one rather than two powers
of the (small) polarization fraction, they will be measured with
better signal-to-noise than polarization autocorrelations.
Indeed, it has been shown \cite{ZSS97,foregpars} that for almost all 
cosmological parameters, the polarization capabilities 
of upcoming high-precision CMB experiments such as MAP and Planck
add information mainly through the cross-polarization signal,
the only exceptions being the reionization and gravity wave parameters.

The goal of the present paper is to present a detailed study of the
cross-polarization from a practical point of view, connecting real-world 
data to physical models. In section \sec{theo}, we argue that the 
dimensionless correlation coefficient $r_\l$ is a more meaningful 
quantity to discuss than the cross power spectrum $C_\l^X$, and 
illustrate how it depends of various cosmological parameters for 
standard adiabatic models.

In \sec{piqcase}, we compute the strongest observational constraints 
to date on $r_\l$ by cross-correlating the polarized Princeton IQU 
Experiment (PIQUE) \cite{H00} with the unpolarized Saskatoon (SK) 
\cite{N97} data set. We do this using the formalism presented in 
\cite{TC01}, which takes into account real--world issues such as 
incomplete sky coverage and correlated noise.
Finally, in \sec{boomcase} we assess the prospects for measuring $r_\l$ 
in the near future by analyzing a simulated version of the upcoming 
$\boom$ 2002 experiment. 


\section{POLARIZATION PHENOMENOLOGY}\label{theo}

Whereas most astronomers use the Stokes parameters $Q$ and $U$ 
to describe polarization measurements, the CMB community uses 
two scalar fields $E$ and $B$ that are independent of 
how the coordinate system is oriented, and are
related to the tensor 
field $(Q,U)$ by a non--local transformation \cite{K97,ZS97,Z98}.
Scalar CMB fluctuations have been shown to generate 
only $E$-fluctuations, whereas gravity waves, CMB lensing and foregrounds
generate both $E$ and $B$. This formalism has been applied to real 
data by the PIQUE \cite{H00} and POLAR \cite{Keating01} teams.


\subsection{The six power spectra}
\label{SixSec}

\begin{figure}[tb]
\preskip
\centerline{\epsfxsize=7.5cm\epsffile{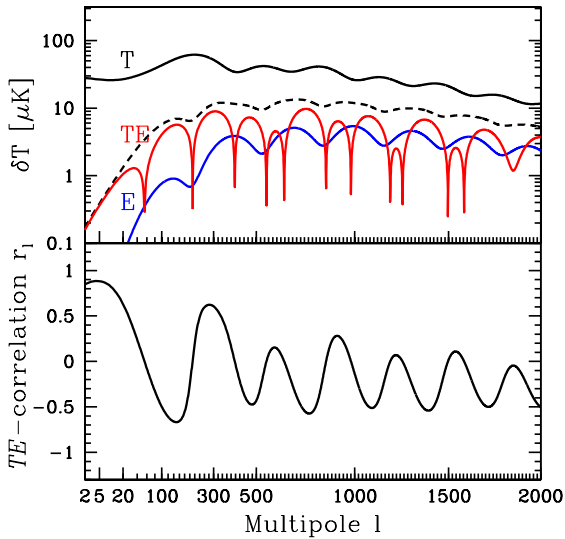}}
\postskip
\caption{\label{bestfitFig}\footnotesize%
	The $T$, $X$ and $E$ power spectra are shown (top panel) for the 
	concordance model of \protect\cite{X01}. 
	The $TE$ correlation coefficient $r^X_\l$ (bottom panel) is the 
	ratio between $C_\l^X$ and the geometric mean of $C_\l^T$ and 
	$C_\l^E$, so in the top panel, $|r^X_\l|$ is just the 
	distance between $TE$ and the geometric average of the $T$ 
	and $E$ curves (dashed curve). 
        }
\end{figure}

Since CMB measurements can be decomposed\footnote{
	Since the transformation between $(Q,U)$ and $(E,B)$ is 
	non-local, the $E$/$B$-decomposition is straightforward only 
	for the case of complete sky coverage. Methods for proceeding 
	in practice for the real-world case of incomplete sky coverage 
	are discussed in \cite{TC01,Z98,Jaffe00,Zalda01,Lewis02,Bunn01}, 
	and we will use such a method  for our real-world calculations 
	below.
        }
into three maps ($T$,$E$,$B$), where $T$ denotes the unpolarized 
component, there are a total of 6 angular power spectra that can 
be measured. We will denote these 
	$C_\l^T$, 
	$C_\l^E$, 
	$C_\l^B$, 
	$C_\l^X$, 
	$C_\l^Y$ and
	$C_\l^Z$, 
corresponding to 
	$TT$,  
	$EE$,  
	$BB$,  
	$TE$,  
	$TB$ and  
	$EB$  
correlations\footnote{
	From here on, we adopt 
	$TT \equiv T$, 
	$EE \equiv E$,
	$BB \equiv B$, 
	$TE \equiv X$, 
	$TB \equiv Y$,
	$EB \equiv Z$, 
	},
respectively. By parity, $C_\l^Y=C_\l^Z=0$ for scalar CMB fluctuations, 
but it is nonetheless worthwhile to measure these power spectra as probes 
of both exotic physics \cite{KK99,XK99,Kamionkowski00} 
and foreground contamination. $C_\l^B=0$ for scalar CMB fluctuations to
first order in perturbation theory \cite{K97,ZS97,Z98,HuWhite97} --- secondary 
effects such as gravitational lensing can create $B$ polarization even if 
there are only density perturbations present \cite{ZSLENS}. The remaining 
three power spectra are plotted in \fig{bestfitFig} (top) for the 
``concordance'' model of \cite{X01} (that of \cite{Efstathiou01} is very 
similar), showing that $C_\l^E$ is typically a couple of orders of magnitude
below $C_\l^T$ on small scales and approaches zero on the very largest
scales (in the absence of reionization).


\begin{figure}[tb]
\preskip
\centerline{\epsfxsize=7.5cm\epsffile{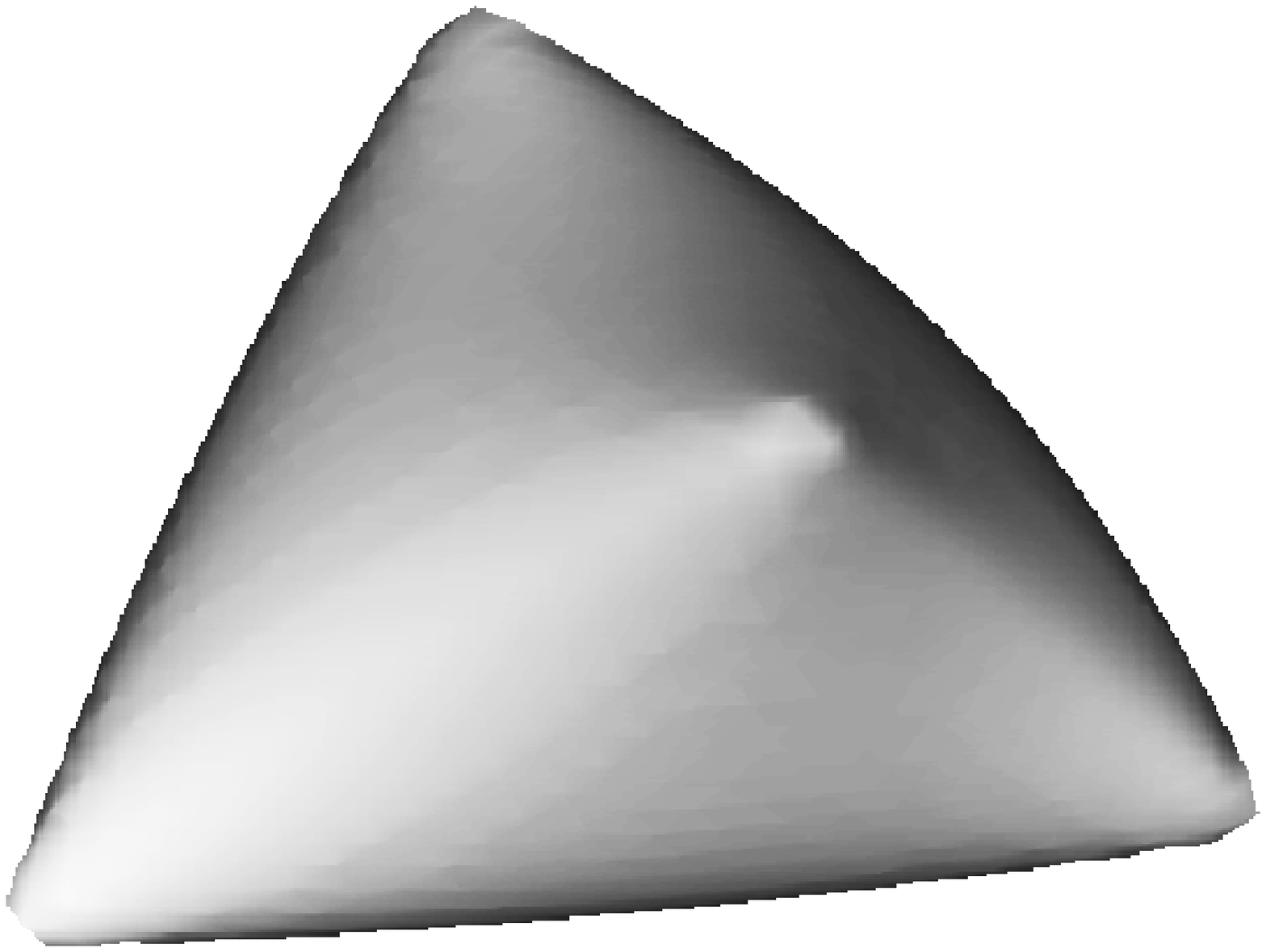}}
\vskip-5.8cm
\centerline{\epsfxsize=7.5cm\epsffile{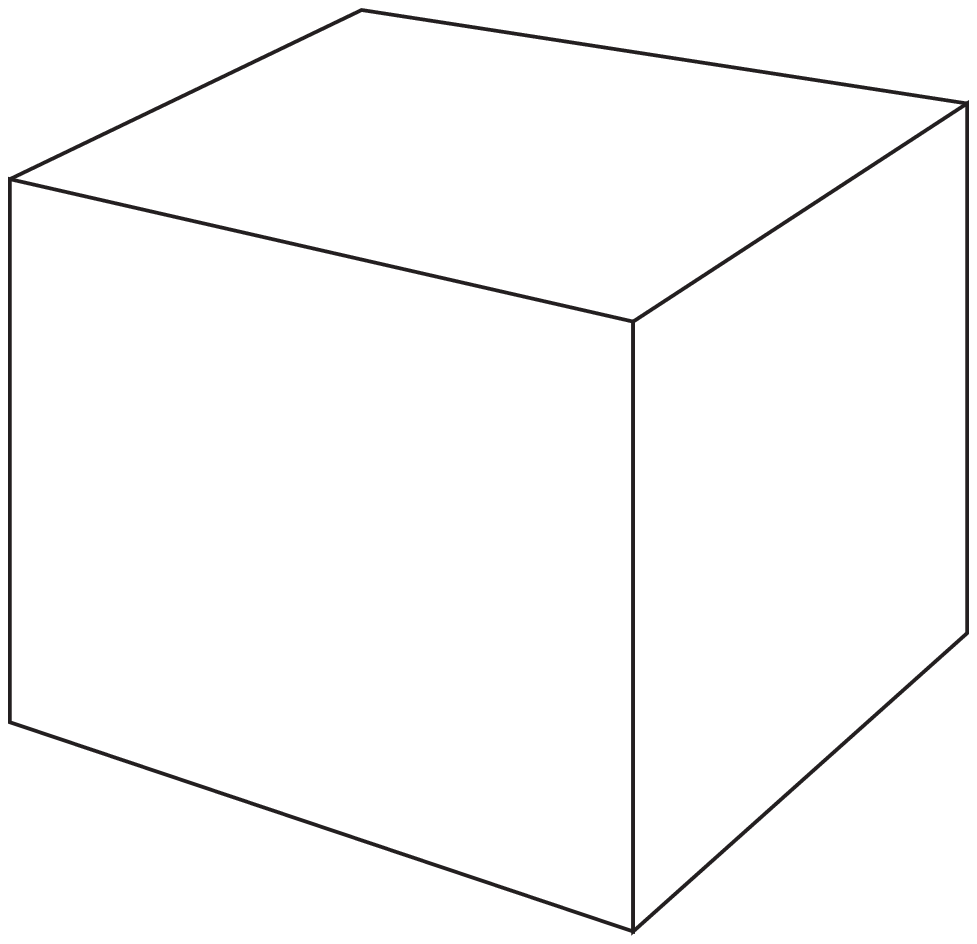}}
\postskip
\caption{\label{tetraFig}\footnotesize%
	 The region of the parameter space $(r^X,r^Y,r^Z)$ allowed by the 
	 generalized Schwarz inequality. The cube extends from $-1$ to $1$ 
	 in all three dimensions. The four corners $(1,1,1)$, $(1,-1,-1)$, 
	 $(-1,1,-1)$ and $(-1,-1,1)$ correspond to $T$, $E$ and $B$ being 
	 perfectly correlated and anticorrelated.
	 }
\end{figure}

\subsection{Covariance versus correlation}

The cross-power spectrum $C_\l^X$ is not well suited for such a logarithmic 
plot, since it is negative for about half of all $\l$-values. A more 
convenient quantity is the dimensionless correlation coefficient $r^X_\l$ 
plotted in \fig{bestfitFig} (lower panel), defined as
\beq{rDefEq}
	r^X_\l\equiv {C_\l^X\over (C_\l^T C_\l^E)^{1/2}},
\eeq
since the Schwarz inequality restricts it to lie in the range 
\beq{SchwarzEq}
	-1\le r^X_\l\le 1.
\eeq
These dimensionless correlations/anticorrelations are seen to be quite strong 
in the sense of being near these limiting values on many scales. For instance, 
if one were to smooth the $T$- and $E$-maps to contain only large angular scales 
$\l\simlt 10$ where $r^X_\l\sim 0.9$, they would be so strongly correlated that
most hot and cold spots would tend to line up\footnote{
	Such a correlation, however, would be very difficult to detect 
	given how small $E$ is expected to be in that $\l$-range}. 
Conversely, if one were to band-pass filter the two maps on scales 
$100\simlt\l\simlt 200$ where $r^X_\l\simlt -0.6$, hot spots in the $T$-map 
would tend to line up with cold spots in the $E$-map.

 More generally, let us also define 
	$r^Y_\l$ $\equiv$ $C_\l^Y/(C_\l^T C_\l^B)^{1/2}$,
	$r^Z_\l$ $\equiv$ $C_\l^Z/(C_\l^E C_\l^B)^{1/2}$. 
Expanding the $T$, $E$ and $B$ maps in spherical harmonics with coefficients 
$a_{\l m}^T$, $a_{\l m}^E$ and $a_{\l m}^B$, the three cross power spectra 
are by definition the covariances
	$C_\l^X=\expec{a_{\l m}^{T*} a_{\l m}^E}$,
	$C_\l^Y=\expec{a_{\l m}^{T*} a_{\l m}^B}$,
	$C_\l^Z=\expec{a_{\l m}^{E*} a_{\l m}^B}$,
so $r^X$, $r^Y$ and $r^Z$ are simply the correlation coefficients between 
$a_{\l m}^T$, $a_{\l m}^E$ and $a_{\l m}^B$.

What values are these three correlation coefficients allowed to take? 
They are real-valued just as $T$, $E$ and $B$, which is most easily seen 
using real-valued spherical harmonics.
The dimensionless correlation matrix corresponding to the vector 
	$(a_{\l m}^T,a_{\l m}^E,a_{\l m}^B)$ is
\beq{RdefEq}
\R=\left(\bs\begin{tabular}{ccc}
1		&$r^X_\l$	&$r^Y_\l$\\
$r^X_\l$	&1		&$r^Z_\l$\\
$r^Y_\l$	&$r^Z_\l$	&1       \\
\end{tabular}\bs\right),
\eeq
and by virtue of being a correlation matrix, it cannot have any negative 
eigenvalues. This implies not only that 
	$|r^X_\l|\le 1$, 
	$|r^Y_\l|\le 1$ and
	$|r^Z_\l|\le 1$, 
but also that the determinant of $\R$ must be non-negative, \ie,
that 
\beq{rConstraintEq1}
    (r^X_\l)^2 + (r^Y_\l)^2 + (r^Z_\l)^2 - 2 r^X_\l r^Y_\l r^Z_\l \le 1
\eeq
for all $\l$. This allowed region in the 3-dimensional space $(r^X,r^Y,r^Z)$ 
is plotted in \fig{tetraFig}, and is seen to resemble a deformed tetrahedron. 
If two coefficients are strongly correlated with each other, 
then they must both have roughly the same correlations with the third,
approaching the limiting case $(r^X,r^Y,r^Z)=(1,1,1)$ (upper right corner).
The other three corners correspond to the allowed possibilities
$(1,-1,-1)$, $(-1,1,-1)$ and $(-1,-1,1)$.

From here on we use $r_\l$ as shorthand for $r^X_\l$. 


\subsection{Cosmological parameter dependence of polarization power spectra}
\label{ParameterSec}

A detailed review of how CMB polarization reflects underlying physical 
processes in given in \cite{HW97}. In this subsection, we briefly review 
this topic from a more phenomenological point of view (see also 
\cite{Jaffe00}), focusing on how different cosmological parameters 
affect various features in the $E$, $B$ and $X$ power spectra and aiming to 
familiarize the reader with, in particular, the correlation spectrum $r_\l$ 
and its cosmology dependence. For more details, the reader is refererred to 
the polarization movies at
 	{\it www.hep.upenn.edu/$\sim$angelica/polarization.html}.

Let us consider adiabatic inflationary models specified by the following 10 
parameters:
	the reionization optical depth $\tau$, 
	the primordial amplitudes $\As$, $\At$ and 
	tilts $\ns$, $\nt$ of scalar and tensor fluctuations, 
and five parameters specifying the cosmic matter budget. The various 
contributions $\Omega_i$ to critical density are for
	curvature $\Ok$, 
	vacuum energy $\Ol$, 
	cold dark matter $\Oc$, 
	hot dark matter (neutrinos) $\On$ and 
	baryons $\Ob$.
The quantities $\ob\equiv h^2\Ob$ and $\od\equiv h^2\Od$ correspond to the 
physical densities of baryons and total (cold + hot) dark matter 
($\Od\equiv\Oc+\On$), and $\fn\equiv\On/\Od$ is the fraction of the dark 
matter that is hot. The baseline values of the parameters here and in the 
movies are for the concordance model of \cite{X01}, $\tau=\Ok=\At=\fn=0$, 
$\Ol=0.66$, $\od=0.12$, $\ob=0.02$, $\ns=0.93$, providing a good fit to current
data from the CMB, galaxy and Lyman Alpha Forest clustering and Big Bang 
nucleosynthesis.


\subsubsection{Polarized versus unpolarized}

If recombination were instantaneous, there would be no polarization at all!

Both the $E$ and the $T$ power spectra carry information about the $z\simgt 10^3$ 
pre-recombination epoch in the form of acoustic oscillations. From a practical 
point of view, there are two obvious differences between the $E$ and $T$ power 
spectra as illustrated by \fig{bestfitFig}:
\begin{itemize}
\item The $E$ power is smaller since the polarization percentage is small, 
      making measurements more challenging. This is because polarization 
      is only generated when locally anisotropic radiation scatters off of free electrons,
      and this only occurs during the brief period when recombination is taking place:
      before recombination, radiation is quite isotropic and after recombination
      there is almost no scattering.
\item Aside from reionization effects, the $E$ power approaches zero on scales 
      much larger than those of the first acoustic peak. 
      This is because the polarization anisotropies are only generated on
      scales of order  the mean free path at recombination and below.
\end{itemize}
As detailed below, changing the cosmological parameters affects the polarized 
and unpolarized power spectra rather similarly except for the cases of 
reionization and gravity waves. All power spectra were computed with the 
CMBfast software \cite{SZ96}.
 

\subsubsection{Reionization}

\begin{figure}[tb]
\preskip
\centerline{\epsfxsize=7.5cm\epsffile{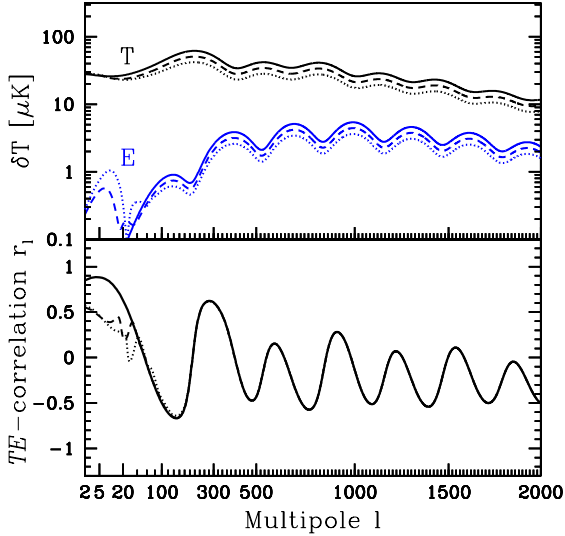}}
\postskip
\caption{\label{tauFig}\footnotesize%
	How the reionization optical depth $\tau$ affects the $T$ and $E$ power 
	spectra (top) and the $TE$ correlation $r_\l$ (bottom). Solid, dashed and dotted 
	curves correspond to for $\tau$=0, 0.2 and 0.4, respectively.
        }
\end{figure}

Reionization at redshift $z_*$ introduces a new scale $\l_*\sim 20 (z_*/10)^{1/2}$ 
corresponding to the horizon size at the time. Primary (from $z\simgt 1000$) 
fluctuations $\delta T_\l$ on scales $\l\gg\l_*$ get suppressed by a factor 
$e^\tau$ and new series of peaks\footnote{These new peaks are caused not by 
       acoustic oscillations, but by a projection effect: they are 
       peaks in the Bessel function that accounts for free streaming, converting 
       local monopoles at recombination to local quadrupoles at reionization.
       }
are generated starting at the scale $\l_*$. \Fig{tauFig} illustrates that although 
these new peaks are almost undetectable in $T$, drowning in sample variance from the 
unpolarized Sachs-Wolfe effect, the are clearly visible in 

\noindent
$E$ since the Sachs-Wolfe nuisance is unpolarized and absent. The models in \fig{tauFig} 
have abrupt reionization giving $\tau\propto z_*^{3/2}$, so higher $z_*$ is seen to 
shift the new peaks both up and to the right.

On small scales, reionization leaves the correlation $r_\l$ unchanged since 
$C_\l^T$ and $C_\l^E$ are merely rescaled. On very large scales, $r_\l$ drops 
since the new polarized signal is uncorrelated with the old unpolarized 
Sachs-Wolfe signal. On intermediate scales $\l\simgt\l_*\sim 20$, oscillatory 
correlation behavior is revealed for the new peaks.

For more details about CMB polarization and reionization see \cite{Z97}.


\subsubsection{Primordial perturbations}

\begin{figure}[tb]
\preskip
\centerline{\epsfxsize=7.5cm\epsffile{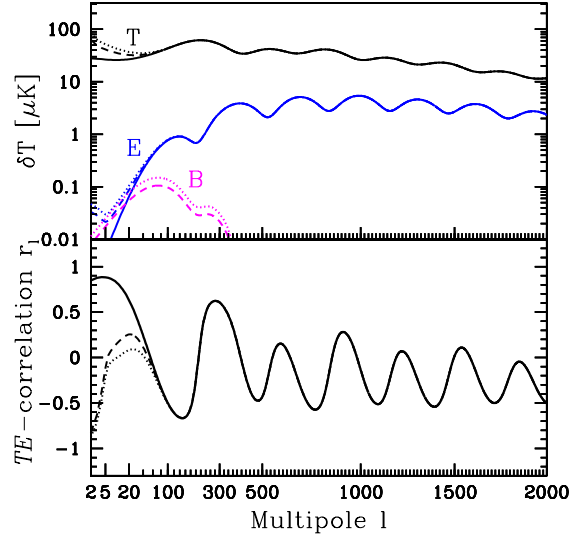}}
\postskip
\caption{\label{AtFig}\footnotesize%
	 How the gravity wave amplitude $\At$ affects the $T$, $E$ and $B$ 
	 power spectra (top) and the $TE$ correlation $r_\l$ (bottom). Solid, dashed
	 and dotted curves correspond to for $\At$=0, 0.2 and 0.4, respectively.
         }
\end{figure}

As seen in \Fig{AtFig}, gravity waves (a.k.a. tensor fluctuations) contribute 
only to fairly large angular scales, producing $E$ and $B$ polarization. 
Just as for the reionization case, unpolarized fluctuations are also produced but 
are difficult to detect since they get swamped by the Sachs-Wolfe effect. As has 
been frequently pointed out in the literature, no other physical effects (except CMB lensing and 
foregrounds) should produce $B$ polarization, potentially making this a smoking gun 
signal of gravity waves.

Adding a small gravity wave component is seen to suppress the correlation $r_\l$ 
in \Fig{AtFig}, since this component is uncorrelated with the dominant signal 
that was there previously. Indeed, this large-scale correlation suppression
may prove to be a smoking gun signature of gravity waves that is easier to 
observe in practice than the oft-discussed ~$B$-signal. ~This ~$TE$-correlation ~suppression 

\noindent
comes mainly from $E$, not $T$: 
since the tensor polarization has a redder slope than the scalar polarization, 
it can dominate $E$ at low $\l $ even while remaining subdominant in $T$.

The amplitudes $\As$, $\At$ and tilts $\ns$, $\nt$ of primordial scalar and tensor 
fluctuations simply change the amplitudes and slopes of the various power spectra:
B is controlled by $(\At,\nt)$ alone, whereas $T$ and $E$ are affected by $(\As,\ns)$ 
and $(\At,\nt)$ in combination. Note that if there are no gravity waves ($\At=0$), 
then these amplitudes and tilts cancel out, leaving the correlation spectrum 
$r_\l$ independent of both $\As$ and (apart from aliasing effects) $n_s$.


\subsubsection{Spacetime geometry}

\begin{figure}[tb]
\preskip
\centerline{\epsfxsize=7.5cm\epsffile{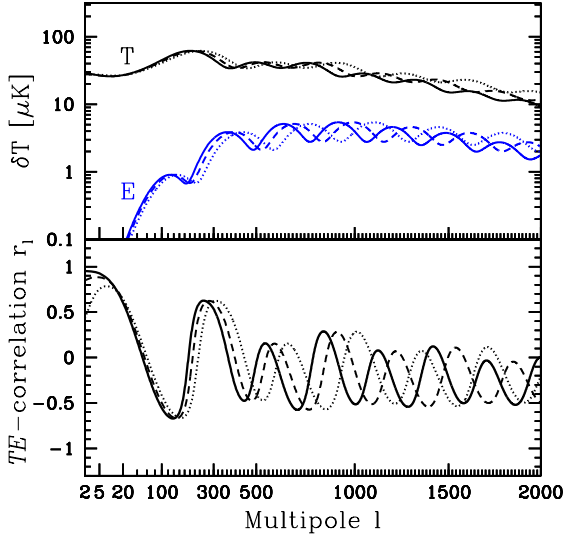}}
\postskip
\caption{\label{OkFig}\footnotesize%
	 How the spatial curvature $\Ok$ affects the $T$ and $E$ power spectra 
	 (top) and the $TE$ correlation $r_\l$ (bottom). Solid, dashed and dotted 
	 curves correspond to $\Ok=-0.1$ (closed model), $0$ (flat model) 
	 and $0.1$ (open model), respectively. Apart from the familiar horizontal 
	 shift, the late ISW effect is seen to suppress the correlation $r_\l$ on 
	 large scales.
         }
\end{figure}

Just as $\As$ and $\ns$, the spacetime geometry parameters $\Ok$ and $\Ol$ affect 
the polarized and unpolarized power spectra in similar ways. $\Ok$ and $\Ol$
were completely irrelevant at $z>10^3$, when the acoustic oscillations were created,
since $\Ok\approx\Ol\approx 0$ at that time regardless of their present values.
As is well-known and illustrated by the above-mentioned movies, the acoustic peak 
features are therefore independent of these parameters, merely shifting sideways 
on a logarithmic plot as geometric effects magnify/shrink the scale of primordial 
fluctuation patterns. On large scales, the late ISW effect creates additional power 
that is completely unpolarized, since this is a pure gravity effect involving no
Thomson scattering. Since the ISW effect is uncorrelated with the primary large-scale 
fluctuations, it suppresses $r_\l$ on large scales as $\Ok$ or $\Ol$ shift away from 
zero as seen in \fig{OkFig}.

\subsubsection{Matter budget}

\begin{figure}[tb]
\preskip
\centerline{\epsfxsize=7.5cm\epsffile{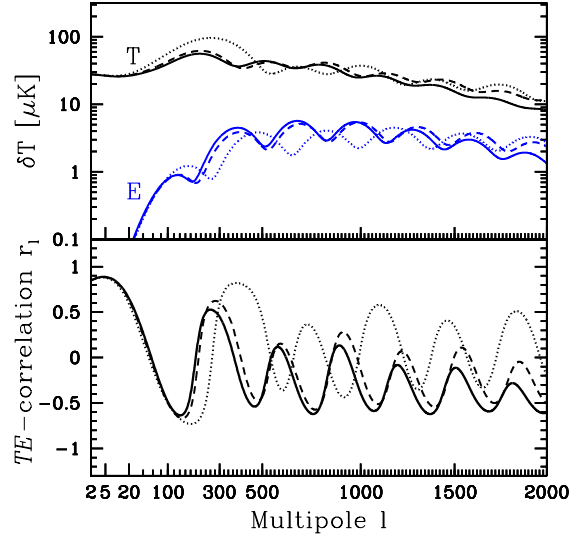}}
\postskip
\caption{\label{baryonFig}\footnotesize%
	 How the baryon fraction $\Omega_b/(\Omega_b+\Omega_d)$ affects the $T$ and 
	 $E$ power spectra (top) and the $TE$ correlation $r_\l$ (bottom). Solid, dashed 
	 and dotted curves correspond to baryon fractions of 0.01, 0.02 and 0.08, 
	 respectively.
         }
\end{figure}

The Primordial CMB signal is dominated by 
fluctuations in the gravitational potential and the density at 
$z\sim 10^3$, whereas the $E$-signal is dominated by 
peculiar velocities on the last scattering surface.
This is why the $E$-spectrum is seen to be out of phase with the 
T-spectrum, peaks in one matching troughs in the other. 
This also explains why increasing the baryon fraction 
$f_b\equiv\Ob/(\Ob+\Od)$ as in \fig{baryonFig} lowers
the polarized peaks, in contrast to the boosting of odd peaks
for the unpolarized case: more baryons lower the sound speed in
the photon-baryon plasma, producing lower velocities.

The remaining matter budget parameters, the cold and hot dark
matter densities, effect $E$ polarization in much the same way $T$.
Increasing the dark matter density $h^2\Od$ shifts the peaks down
and to the left.
The $T$, $E$ and $r_\ell$ power spectra change hardly at all 
when changing the fraction $\On/\Od$ of the dark matter that
is hot. This is because the neutrinos were already quite cold (nonrelativistic) 
at the time the CMB fluctuations are formed.


\section{Case study I: {\rm PIQUE} }\label{piqcase}

We now turn to the issue of measuring cross-polarization in practice, 
including issues of methodology, window functions and leakage.
We analyze an existing data set in this section, then turn to simulations 
of upcoming data in  \sec{boomcase} .


\subsection{Data}

\begin{figure}[tb]
\preskip
\centerline{\epsfxsize=8.0cm\epsffile{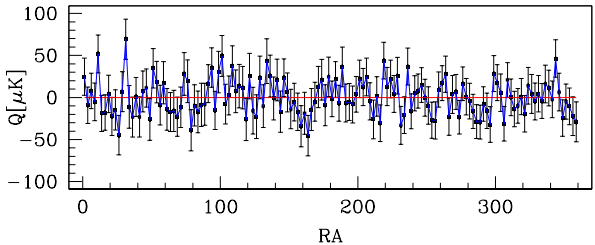}}
\vskip-4.3cm
\caption{\label{PIQmapFig}\footnotesize%
	 The deconvolved PIQUE data consists of the Stokes $Q$-parameter 
	 as a function of position along a $1^\circ$ radius circle around 
	 the North Celestial Pole. One quarter of the modes are projected 
	 out as described in the text.
	 }
\end{figure}

PIQUE was a CMB polarization experiment on the roof of the physics 
building at Princeton University. It used a single 90 GHz correlation 
polarimeter with FWHM angular resolution of $0^{\circ}.235$, and observed 
a single Stokes parameter $Q$ in a ring of radius of $1^\circ$ around the 
North Celestial Pole (NCP) \cite{H00} (see \fig{PIQmapFig}). 

During one day, the telescope was able to observe on the order of 20 
independent points on this ring, chopping slowly (every few seconds) 
between two points separated by 90$^{\circ}$ along this circle. The 
polarized sky signals detected at these two points had opposite signs 
($\pm Q$) and were six hours out of phase\footnote{
        See \cite{H00} and $dicke.princeton.edu$ for more details.
	}.
This means that the experiment did not measure individual $Q$-values, 
but sums of two. To simplify subsequent calculations, we eliminated this 
complication using the deconvolution method described in appendix D of 
   \cite{X02} to recover a filtered version of the $Q$-map. 
Specifically, the ring was pixelized into 144 angular bins, and we denoted
the corresponding $Q$-values $Q_1,...,Q_{144}$. Let us focus on four pixels 
forming a perfect square in the sky, say  pixels 1, 37, 73 and 109, and 
group them into a vector $\x$. Because of the chopping, PIQUE measured
not $\x$ but the linear transformation $\A\x$, where

\bigskip
\beq{Aeq}
    \A =
    \left(\bs\begin{tabular}{cccc}
	 1&1&0&0\\
	 0&1&1&0\\
	 0&0&1&1\\
	 1&0&0&1\
\end{tabular}\bs\right) ~{\rm and}
\quad
    \x =
    \left(\bs\begin{tabular}{c}
	 $Q_{1}  $\\
	 $Q_{37} $\\
	 $Q_{73} $\\
	 $Q_{109}$
\end{tabular}\bs\right).
\eeq
The matrix $\A$ is singular, with a vanishing eigenvalue corresponding
to the vector $(1,-1,1,-1)$. Our deconvolution therefore sets this 
particular mode to zero, and encodes our lack of information about its
value as near-infinite noise for this mode in the map noise covariance 
matrix. Since there are a total of 36 such pixel quadruplets, our 
deconvolved data set, which is plotted in \fig{PIQmapFig}, thus has
36 such unmeasured modes. 

Because PIQUE was insensitive to the unpolarized CMB component, we 
cross-correlated the $Q$-data from PIQUE with $T$-data from the SK map 
\cite{T97}, deconvolved and pixelized as described in \cite{X02}. 
To take advantage of cross-polarization information from spatially 
separated pixels, we used SK pixels not merely from the PIQUE circle, 
but from a filled disk of radius $3^\circ$ around the NCP, a total 
of 288 $0.31^\circ\times 0.31^\circ$ pixels. We found that further 
increasing the size of the SK disk did not significantly tighten our 
constraints. Our final data vector combines the PIQUE $Q$-data and the 
SK $T$-data and thus contains $144+288=432$ pixels.


\subsection{Method}

We compute the six power spectra describe in \sec{SixSec} using quadratic 
estimator method as described in \cite{TC01}, computing fiducial 
power spectra with the CMBfast software \cite{SZ96} using cosmological 
parameters from the concordance model from \cite{X01}. We also perform a likelihood
analysis as described below.

A key challenge is separating the six power spectra ($T, E, B, X, Y, Z$). 
A generic quadratic band power estimator (a quadratic combination of $T$, 
$Q$ and $U$ pixels) will probe a weighted average of all six power spectra, so 
measurements of the six can in principle be afflicted by as many as 
$\left({6\atop 2}\right)=15$ types of unwanted ``leakage'' whereby a measurement of 
one power spectrum picks up contributions from another. In \cite{TC01} it was 
argued that susceptibility to systematic errors could be reduced by chosing the 
``priors'' that determine the quadratic estimator method to have
vanishing cross-polarizations, $X=Y=Z=0$, and it was shown that this simplification
came at the price of a very small (percent level) increase in error bars.
In Appendix A, we show that this choice has an important added benefit: 
exploiting a parity symmetry, it eliminates 14 out of the 15 leakeges, with only 
the much discussed \cite{TC01,Z98,Jaffe00,Zalda01,Lewis02,Bunn01} $E-B$ leakage remaining.
Note that whereas parity-conservation {\it in the laws of physics} predicts 
$Y=Z=0$, the proof given in Appendix A is purely  geometric in nature, and holds 
even if parity-violating physics, foregrounds, {\etc}, produce non-vanishing $Y$ and $Z$.

This result is useful in all but eliminating the leakage headache, which would otherwise
complicate both calculation and interpretation. It also has important implications for 
other approaches, notably the currently popular maximum-likelihood (ML) method as 
implemented by the MADCAP software \cite{borrill} and applied to Maxima, \boom, DASI 
and other experiments: generalizing it to polarization in the obvious way will produce 
the exactly the sort of leakage that our method eliminates. However, this problem 
can be eliminated as described below.

The quadratic estimator (QE) method is closely related to the ML method: the latter 
is simply the quadratic estimator method with $\B=\F^{-1}$ in \eq{QdefEq}, iterated so that the 
fiducial (``prior'') power spectrum equals the measured one \cite{BJK}.
For a detailed comparison between these two methods, see  section IV.B. in \cite{TC01}.
The ML method has the advantage of not requiring any prior to be assumed.
The QE method has the advantage of being more accurate for constraining 
cosmological models --- since it is quadratic rather than highly non-linear,
the statistical properties the measured band power vector $\q$ can
be computed analytically rather than approximated, which allows the likelihood 
function to be computed directly from $\q$ (as opposed to $\x$), in terms of 
generalized $\chi^2$-distributions \cite{Wandelt}.

Both methods are unbiased, but they may differ as regards error bars.
The QE method can produce inaccurate error bars if the prior is
inconsistent with the actual measurement.
The ML method Fisher matrix can produce inaccurate error bar estimates
if the measured power spectra have substantial scatter due to noise or
sample variance, in which case they are unlikely to describe the
smoother true spectra.
A good compromise is therefore to iterate the QE method once and choose
the second prior to be a rather smooth model consistent with
the original measurement. 
The lesson to take away from Appendix A is that if the ML method is used,
one should reset $X=Y=Z=0$ before each step in the iteration, thereby eliminating
all leakage except between $E$ and $B$.


\bigskip
\bigskip
{\bf \centerline{Table 1 -- Polarization power spectrum$^{(a)}$}} 
\bigskip
\centerline{  
\begin{tabular}{cccc}
\hline
\multicolumn{1}{l}{}			            &
\multicolumn{1}{c}{$\dT ^2\pm \sigma^2\> [\mu K^2]$}&
\multicolumn{1}{c}{$\dl \pm \sigl$}	            & 
\multicolumn{1}{r}{$\dT\> [\mu K]^{(b)}$}           \\
\hline
\hline
 $T$   &  2183  $\pm$905      	 &151.3$\pm$ 85.8	&46.7$^{+9}_{-11}$ \\
 $E$   &    -2.4$\pm$ 17.4    	 &153.4$\pm$ 87.6	&$<$~3.9~(5.7)     \\
 $B$   &    -8.1$\pm$ 16.0    	 &141.6$\pm$ 73.7	&$<$~2.8~(4.9)     \\
 $X$   &  -145.5$\pm$222.2    	 &137.4$\pm$ 69.4	&$<$~8.6(17.3)     \\
 $Y$   &    -7.9$\pm$206.2    	 &151.3$\pm$ 82.5	&$<$14.1(20.1)     \\
\hline
\end{tabular} 
}
\vskip 0.1cm
\noindent{\small $^{(a)}$Results from combined PIQUE and SK data;} \\ 
\noindent{\small $^{(b)}$Values in parentheses are 2-$\sigma$ upper limits.} 
\bigskip

\subsection{Results}

Table 1 shows the result of our band-power estimation. Here we use 50 multipole 
bands of width $\Delta\l=20$ for each of the six polarization types 
$(T,E,B,X,Y,Z)$, thereby going out to $\l=1000$, and average the measurements 
together into a single number for each polarization type to reduce noise.
We eliminate sensitivity to offsets by projecting out the mean (monopole) from 
the $T$ and $Q$ maps separately. The values shown in parentheses in the most
right column of Table 1 are 2-$\sigma$ upper limits (those are also the values 
that we present here as our upper limits).

The detection of unpolarized power is seen to be consistent with that published
for the full SK map \cite{N97}. The table shows that we detect no polarization or 
cross-polarization of any type, obtaining upper limits, just as the 
concordance model predicts. No results of $Z$ are reported since PIQUE provides 
only $Q$ data (note that $Q$ and $U$ are needed to isolate $Z$-polarization --- 
we will include $Z$ in in \sec{boomcase}).
 
The window functions reveal substantial leakage between $E$  and $B$, so the limits 
effectively constrain the average of these two spectra rather than both separately.
For this reason, and to recast our constraints in terms of the $TE$ correlation coefficient 
$r_\ell$, we complement our band-power analysis with a likelihood analysis where 
we assume $B=0$. Specifically, we set $B=Y=Z=0$ and take each of the remaining power 
spectra  $(T,E,X)$ to be constant out to $\l=1000$. 

\begin{figure}[tb]
\preskip
\centerline{\epsfxsize=8.5cm\epsfysize=9.5cm\epsffile{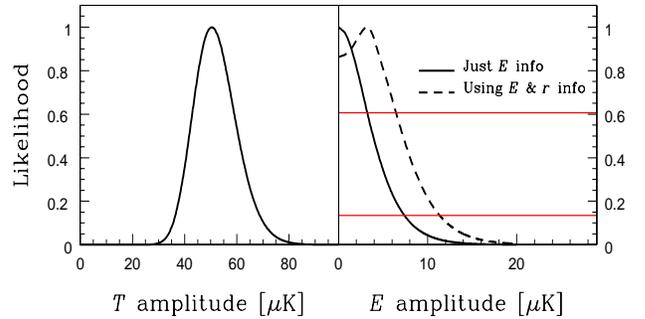}}
\vskip-4.6cm
\caption{\label{LikeTE}\footnotesize%
	Likelihood results using PIQUE $Q$-information alone (right panel, solid line)
        and using both PIQUE $Q$- and SK $T$-information and marginalizing
        (remaining two curves). From top to bottom, the two horizontal red 
	lines correspond to 68\% and 95\% of C.L., respectively.
	}
\end{figure}

We first perform a simple 1-dimensional likelihood analysis for the parameter $E$ using 
the PIQUE data alone (discarding the SK information), obtaining 
the likelihood function (bottom solid line in right 
panel of \fig{LikeTE}) in good agreement with that published by the PIQUE team \cite{H00}. 
They find 95\% of the area for $E<10\mu$K --- our likelihood curve drops by a factor $e^{-2}$ 
at a slightly lower value 
$E\approx 7\mu$K as expected, since the likelihood curve is asymmetric and hence
highly non-Gaussian. This non-Gaussianity also 
means that the precise confidence levels of the upper limits in Table 1 should 
be taken with a grain of salt. The $B$-limit is rather low because noise fluctuations
give a negative best estimate (recall that what is measured is the 
sky power minus the expected noise power, which can be negative), and 
this preference for negative $B$ pulls down the $E$ estimate too since the
strong leakage implies that it is really measuring a weighted average of $E$ and $B$.

\begin{figure}[tb]
\preskip
\centerline{\epsfxsize=8.0cm\epsffile{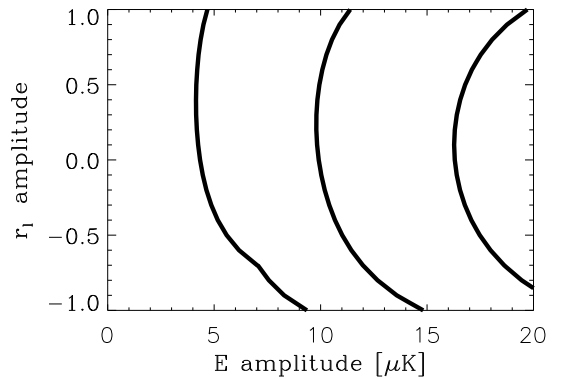}}
\postskip
\caption{\label{Er}\footnotesize%
	Joint constraints on $E$ polarization and $r_\ell$ after
	marginalizing over $T$. From left to right, the contours show that the 
	likelihood function has dropped to $e^{-1.1}$, $e^{-3.0}$ and $e^{-4.6}$ 
	times its maximum value, which would correspond to 68\%, 95\% and 99\% 
	limits if the likelihood were Gaussian. 
	For comparison, the concordance model predicts ($E,r_\ell$)=(0.82,-0.65) 
	at $\l$=137, the center of our window function for $X$ (see Table 1).
        }
\end{figure}

We then compute the likelihood function including both PIQUE and 
SK data in the 3-dimensional space spanned by $(T,E,r_\ell)$ and compute constraints 
on individual parameters or pairs by marginalizing as in \cite{X01}. 
\Fig{LikeTE} shows that this produces a $T$-measurement $T\approx 50\mu$K, consistent 
with that for the full SK map \cite{N97} (left panel).
 \Fig{Er} shows our constraints in the $(E,r_\ell)$-plane after marginalizing over $T$.
The ``C'' shape of the contours in here are not generic: we performed a series of 
Monte Carlo simulations, and some produced ``backward C'' shapes instead, others 
``S''-like shapes. This figure also shows that our constraints on the cross-polarization 
are weaker than the Schwarz inequality $r_\ell \le 1$, so in this 

\begin{figure}[tb]
\preskip
\centerline{\epsfxsize=8.0cm\epsffile{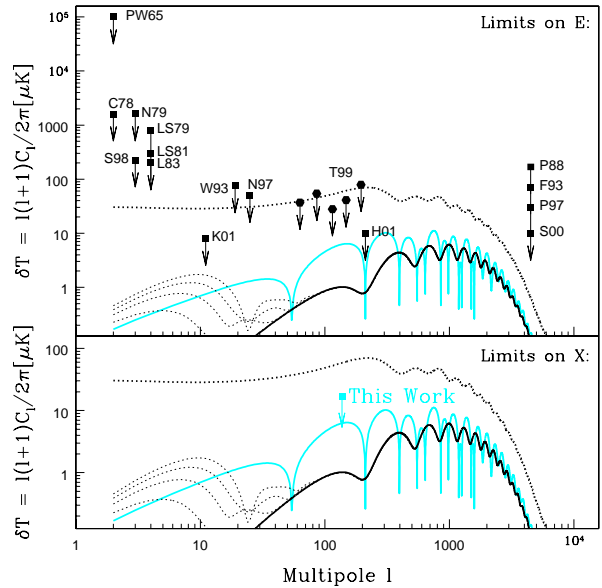}}
\postskip
\caption{\label{summaryFig}\footnotesize%
 	Summary of upper limits on polarization so far. From top to bottom,
	the three curves show the concordance model predictions for 
         $C_\l^T$, $C_\l^E$ and $C_\l^X$,	respectively.
         Four reionization models with $\tau$=0.1, 0.2, 0.3  
	and 0.4 are also plotted (left thin lines from bottom to top).
	The limits are: 
	       PW65\protect\cite{PW65},
	        C78\protect\cite{C78},
	        N79\protect\cite{N79},
	       LS79\protect\cite{LS79},
	       LS81\protect\cite{LS81},
	        S98\protect\cite{S98},
	        L83\protect\cite{L83},
	        W93\protect\cite{W93},
	        N97\protect\cite{N97},
	        T99 {\it hexagons}\protect\cite{T99},
	        P88\protect\cite{P88},
	        F93\protect\cite{F93},
	        P97\protect\cite{P97},
	        S00\protect\cite{S00},
	        H01\protect\cite{H00} and
	        K01\protect\cite{Keating01} (all in upper panel) and
	       our new upper limit on $X$ (lower panel).
 	}
\end{figure}

\noindent
sense the data has taught us nothing new. This funny situation is unique to 
measuring correlations, since any measurement of say $T$ or $E$, however noisy, 
will always rule out some class of theoretically 
allowed models. However, it is important to point out that this failure to beat 
the Schwartz inequality does not mean that an experiment is far from detecting 
polarization: as we saw in \sec{ParameterSec}, the actual correlation coefficient 
can be of order unity, so the step from overcoming the Schwartz bound to detecting 
an $X$-signal may in fact be quite small. \Fig{summaryFig} summarizes all polarization 
limits published to date, and indicates that the detection of cross-polarization may 
indeed be just around the corner. 

Cross-polarization information could, in principle, reduce error bars on $E$ as well. 
As an extreme example, if we had measured that $r_\ell \approx 1$ on the angular scales 
probed by PIQUE, with tiny error bars, then we would know the exact spatial template 
of the $E$-map from the $T$-map and could therefore fit for its amplitude quite accurately.
The effect of adding cross-polarization information on the PIQUE $E$-limits is shown 
in \Fig{LikeTE} (dashed line in right panel). In this case, we had no meaningful 
constraints on $r_\ell$, so it is not surprising that this approach does not help us: 
we see that adding the unpolarized temperature simply weakens the upper limit slightly 
courtesy of noise fluctuations. 

Since $r_\l$ is expected to oscillate between positive and negative values,
using a flat (constant) $r_\l$ in the likelihood analysis runs the risk of 
failing to detect a signal that is actually present in the data, 
canceling out positive and negative detections at different angular scales.
This is not likely to have been a problem in our case,
since $r_\l$ is uniformly negative in our sensitivity range $\l=137\pm 69$
for the concordance model, but for future experiments with higher signal-to-noise,
it will be important to parametrize $r_\l$ in a more physical way - either with 
separate bandpowers in multiple bands or directly in terms of cosmological parameters.


\subsection{Future prospects}

\begin{figure}[tb]
\preskip
\centerline{\epsfxsize=8.0cm\epsffile{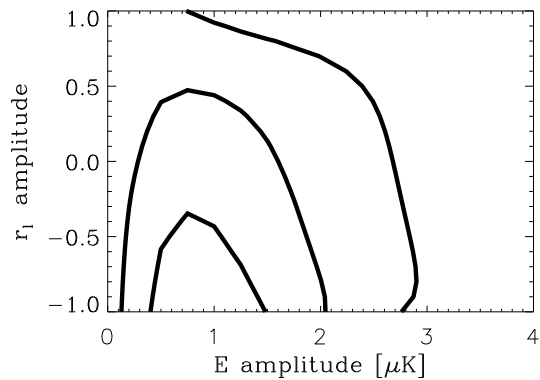}}
\postskip
\caption{\label{Er_tinyT}\footnotesize%
        $E$--$r_\ell$ likelihood for the case of lower noise 
        (50 $\mu$K per SK pixel, 5 $\mu$K  per PIQUE pixel), showing 
        that this sensitivity level can give interesting 
        constraints on $r_\ell$. The data used here is a Monte Carlo simulation
        with ($E,r_\ell$)=(0.82,-0.65), which is what the concordance model predicts 
	at $\l$=137, the center of our window function for $X$ (see Table 1).
	}
\end{figure}

What improvements in data would enhance the scientific potential of an experiment
like PIQUE the most? Reducing the noise level in the polarization measurements 
would obviously improve the limits on $E$, $B$ and $r_\l$. We find that substantial 
reductions would be needed  to make a qualitative difference -- a simulation merely 
doubling the amount of data, adding $U$-polarization, did not produce an 
$r_\l$-detection or beat the Schwarz inequality.
\Fig{Er_tinyT}  shows which sort of sensitivity improvements are needed for a 
marginal detection in the $E$--$r_\l$ plane using the same number and position
of pixels as for PIQUE.
The Schwarz bound inequality is seen to be beaten by a substantial margin on the high side. 
This illustrates the potential value of experimental groups coordinating to observe 
the same sky regions.
Increased sky coverage and a more two-dimensional geometry
clearly helps, and we explore such an example in the next section. 

\bigskip
\bigskip
\bigskip



\begin{figure}[tb]
\preskip
\centerline{\epsfxsize=8.5cm\epsffile{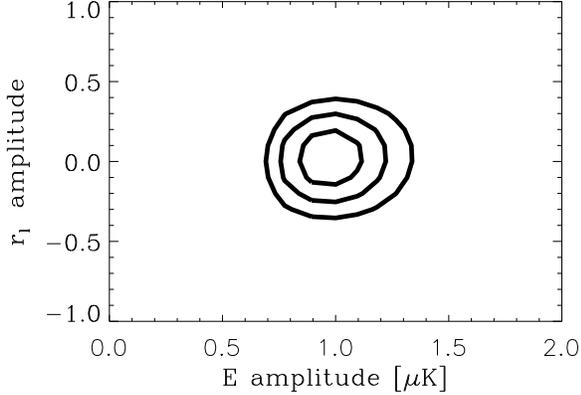}}
\postskip
\caption{\label{boomEr}\footnotesize%
  	Constraints in the $E$-$r_\ell$ plane from a simulation of the $\boom$ experiment.
 	From outside to inside, the likelihood contours are same as in \fig{Er}.   
	A fiducial model with $C_\l^E=1\mu$K and $C_\l^X$=0 was used in this calculation.
 	}
\end{figure}

\section{Case study II: $\boom$ data}\label{boomcase}

We begin case study II by extending the results presented in \cite{TC01}. 
In this section we quantify the ability of $\boom$ to separate the $Y$ and $Z$ 
correlations using $(T,Q,U)$ maps. 
 
We pixelize our sky patch using the equal-area icosahedron method \cite{icosahedron} 
at resolution levels 35, corresponding to 361 $\boom$ pixels\footnote{
   	We use the icosahedron pixelization since it has the roundest (mainly 
	hexagonal) pixels and is highly uniform. Although we did not use it here, 
	the HEALPIX package \cite{HEALPIX} is useful in allowing 
	azimuthal symmetry to be exploited for saving computer time.
	}.
We apply the quadratic estimator described in \cite{TC01} just as we did for PIQUE, 
with the fiducial power spectra $C_\l^T$ computed using CMBfast software \cite{SZ96} 
using cosmological parameters from the concordance model from \cite{X01}. In our 
fiducial model we set $C_\l^E=1\mu$K$^2$ and $C_\l^B=C_\l^X=C_\l^Y=C_\l^Z=0$, and 
eliminate sensitivity by offsets by projecting out the mean (monopole) for $T$, $Q$ 
and $U$ maps separately.

\Fig{boomEr} shows how important the unpolarized counterpart is when constraining 
the polarized power spectrum: high signal--to--noise temperature data is seen to 
substantially improve how well $E$ can be constrained. The $1\mu$K $E$ 
signal used in our case study is detected at high significance when using polarization 
data alone. Although adding $T$-information is obviously necessary to constrain $r_\l$, 
we found that it helped only marginally for constraining $E$: we repeated our likelihood 
analysis using $T$, $Q$ and $U$ data jointly, marginalizing over $T$ and $r_\l$, and 
obtained essentially unchanged error bars. 

\begin{figure}[tb]
\preskip
\centerline{\epsfxsize=8.5cm\epsffile{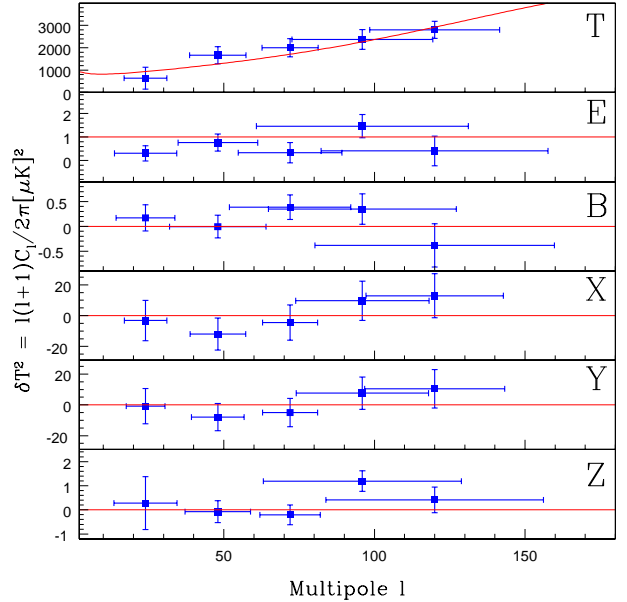}}
\postskip
\caption{\label{boompowerspectrum}\footnotesize%
        $\boom$ power spectrum $T,E,B,X,Y$ and $Z$. The power spectrum was calculated 
 	for 5 bands of size $\dl$=24, $C_\l^T$ was computed using CMBfast software, and 
	we assumed $C_\l^E$=1 and $C_\l^B=C_\l^X=C_\l^Y=C_\l^Z=0$ when running our 
	Monte Carlos.
        }
\end{figure}

Finally, \Fig{boompowerspectrum} shows that we can separate the $Y$ and $Z$ correlations 
if we use $(T,Q,U)$ maps jointly. We also verified numerically that there is no leakage 
between the $TB$ and $EB$ correlations, as proved in \sec{appendix}. 
Note that the results in this section underestimates the true power
of Boomerang 2002 by only using enough pixels to probe the power out to 
$\l\sim 120$, the key intention being simply to demonstrate that the both experimental sensitivity
and our analysis method is adequate. 
When the real data is available, a more computer intensive analysis
with or order $10^4$ pixels will be worthwhile.


\section{CONCLUSIONS}\label{conclusions}

We have presented the first attempt at measuring the CMB cross-polarization, using the 
PIQUE and SK data sets. We obtain upper limits of 
	$TE \equiv X <$  8.6 $\mu$K (95\% CL) 
	and
	$TB \equiv Y <$ 14.1 $\mu$K (95\% CL). 
Limits for $E$ are in concordance with the values presented in \cite{H00}, and
are much higher than the expected NCP foreground levels 
\cite{dOC97,H00}\footnote{ 
		\cite{H00} extrapolated an upper limit of 0.5 $\mu$K for the 
		polarized dust emission from the IRAS 100 $\mu$m map \cite{iras} and 
		an upper limit of 0.4 $\mu$K for the polarized synchrotron emission 
		from the Brouw \& Spoelstra \cite{bs} and Haslam \cite{haslam} maps --- 
		both limits were for the NCP region and the PIQUE observing frequency.
		}.
We also discuss theoretical and practical issues relevant to measuring cross-polarization 
and illustrate them with simulations of the upcoming $\boom$ 2002 experiment:
we find that substantial improvements on measuring the polarized power spectra would be 
possible if the noise could be lowered in the unpolarized map used for the cross-correlation. 
Among other things, we show that the well-known problem of $EB$ leakage, which complicates 
measurements of $E$ and $B$ power, vanishes when measuring the three cross-polarization 
power spectra ($TE$, $TB$ and $EB$).



\section{Appendix}\label{appendix}

In this appendix, we prove the no-leakage theorem described in the text.
Specifically, we show that that our quadratic estimator method gives 
no leakage between any of the 15 power spectrum pairs except E/B.
 
We follow the notation of \cite{TC01} throughout. We let $\x$ denote the vector that
contains the measured temperature and Stokes parameters at each pixel and
consider quadratic estimators of the different power spectra,
\beq{qDefEq}
q_i\equiv\x^t\Q_i\x = \tr[\Q_i\x\x^t],
\eeq
where $i$ labels at the same time both the polarization type $P$
($T,E,B,X,Y$ and $Z$) and the multipole $\l$ to be measured. 
$q_i$ probes a
weighted average of the power spectra,
\beqa{qMeanEq}
\expec{q_i}&=&\tr[\Q_i\C]  \nonumber \\
            &=& b + \sum_{P'=1}^6\sum_{\l'=2}^\lmax W^{\l P}_{\l'P'} C_{\l'}^{P'},
\eeqa
where $b\equiv \tr[\Q_i\NN]$ is the contribution from experimental noise,
$\NN$ is the noise covariance matrix,
$\SS$ is the covariance matrix due to the cosmological signal,
and $\C=\NN+\SS$ is the total covariance matrix. 
In equation (\ref{qMeanEq}) we introduced generalized window functions,
\beq{WindowDefEq}
W^{\l P}_{\l' P'}\equiv\tr[\Q_i\P_{i'}],
\eeq
where $\P=\partial \C /\partial C_l^P$.
For a fixed $(P,\l)$, they show the expected contributions
to the band power estimate $q_i$ not only from different $\l$-values, but also from different
polarization types. For instance, an estimate of $E$-polarization may 
inadvertently pick up a contribution from $B$-polarization as well, since it is difficult
to separate the two with only partial sky coverage \cite{TC01,Z98,Jaffe00,Zalda01,Lewis02,Bunn01}.
Such {\it leakage} between different polarization types is clearly undesirable since it
complicates interpretation of the measurements.
The aim of the appendix is to show that for our particular choice of estimator 
$\Q_i$, most elements of the window functions $W^{\l P}_{\l' P'}$ vanish,
and that there is indeed no other leakage than between any E and B --- not, say,
between X and Y or between E and Z.

In \cite{TC01} it was shown that the quadratic estimator defined by
\beq{QdefEq}
\Q_i = {1\over 2}\norm_i\sum_j B_{ij} \C^{-1}\P_j\C^{-1},
\eeq
distills all the cosmological information from $\x$ into the 
(normally much shorter) vector $\q$ if $\C$ is the true 
covariance matrix. Moreover, if $\C$ is a reasonable 
estimate of the true covariance matrix, then the data compression step 
of going from $\x$ to $\q$ destroys information only to second order.
In \eq{QdefEq}, $\B$ is an arbitrary invertible matrix
and the normalization constants $\norm_i$ are chosen so that 
all window functions sum to unity. 
This means that we can interpret $q_i$ as measuring a weighted
average of our unknown parameters, the window giving the weights.

For our estimates we constructed the covariance matrix appearing in
equation (\ref{QdefEq}) setting $C_l^X=C_l^Y=C_l^Z=0$ and
$C_l^E=C_l^B$. As explained in detail in \cite{TC01}, 
the choice $C_l^X=C_l^Y=0$ was made
because it makes the estimates of the temperature power spectra only
depend on quadratic combinations involving two measured temperatures,
the estimates of $X$ and $Y$  only depend on quadratic combinations
involving  one Stokes parameter and one temperature and the estimate of
the $E$, $B$ and $Z$ spectra only depend on products of two Stokes
parameters (otherwise additional non-intuitive terms get included, say
temperature autocorrelations when measuring $E$, increasing susceptibility
to systematic errors).
These facts  immediately imply that the only mixed windows
that could potentially be non-zero are 
$W^{\l E}_{\l'B},
\ W^{\l X}_{\l' Y},
\ W^{\l E}_{\l' Z},
\ W^{\l B}_{\l' Z}$, 
\ie, that of the $\left({6\atop 2}\right)=15$ types of potential leakage, 
only EB, XY, EZ and BZ leakage is possible.
What we will show in this
appendix using an  argument based on the parity of the different fields
is that only $W^{\l E}_{\l'B}$ is non-zero for our choice of $\C$.

We consider a parity transformation such that the coordinate $\r^\prime=
-\r$, where primes indicate the coordinates after the transformation. We
define the operator $\Pr$ to be the one that transforms the vector of
temperatures and 
Stokes parameters $\x$ under the parity transformation, 
$\x^{\prime}=\Pr \x$. It satisfies $\Pr^2 =\I$, the identity.  For example, if the Stokes parameters at
each point on the sky were defined with respect to the spherical
coordinate system, then $\Pr$ transforms them as $Q^\prime = Q$ and
$U^\prime = -U$. Under a parity  transformation, we also have
$T^\prime=T$, $E^\prime=E$ and $B^\prime=-B$.

To understand how the matrices $\C$ and $\P$ behave under parity, we need
to understand their structure.  Let us consider an arbitrary pair of
points labeled  $A$ and $B$. The matrices are most easily described when
the Stokes parameters are measured with respect to the natural frame,
that is when $Q$ is defined as the differences in intensity in the
directions parallel and perpendicular with respect to the great circle
that joins $A$ and $B$. In that case the structure of the covariance
matrix is,
\beqa{struccov}
\expec{T_A T_B} &=& \sum_l f_1^\l C_\l^T   \nonumber \\
\expec{Q_A Q_B} &=& \sum_l f_2^\l C_\l^E + f_3^\l C_\l^E  \nonumber \\
\expec{U_A U_B} &=& \sum_l f_2^\l C_\l^B + f_3^\l C_\l^B  \nonumber \\
\expec{T_A Q_B} &=& \sum_l f_4^\l C_\l^X   \nonumber \\
\expec{T_A U_B} &=& \sum_l f_4^\l C_\l^Y   \nonumber \\
\expec{U_A Q_B} &=& \sum_l (f_2^\l-f_3^\l) C_\l^Z   
\eeqa
where the coefficients $f^\l_i$ are known  functions of the angular
separation between $A$ and $B$. From the above expressions we can
calculate what the different $\P$-matrices are  in this frame. The
matrix $\P_T$ has non-zero entries only for the terms involving $T_A$
and $T_B$,  $P_E$ and $P_B$ only terms involving $Q_A$ $Q_B$ and $U_A$
$U_B$, $\P_X$ only for those involving $Q_A$ $T_B$ and $Q_B$ $T_A$, 
$\P_Y$ only for those involving $U_A$ $T_B$  and $U_B$ $T_A$ and  $\P_Z$
only for those involving $Q_A$ $U_B$ and $Q_B$ $U_A$. Moreover  if when
constructing $\C$ we took  $C_\l^X$ and $C_\l^Y$  to be zero, $\C$ is
block diagonal, with no terms that mix $T$ and $Q$ or $T$ and $U$.  It
is clear then why our estimators of $X$ and $Y$ only contain $TQ$ and
$TU$ terms.

Under a parity transformation in this coordinate system, the Stokes
parameters transform as $T' =T$ , $Q' =Q$ and $U' =-U$. Under parity in
(\ref{struccov}), the last two equations change sign while the others
remain the same.  Thus if  $\C$ was constructed assuming that all the
cross spectra were zero, which implies that the last three expectation
values in  equation (\ref{struccov}) are zero, then $\Pr \C \Pr = \C$
which implies that $\Pr \C^{-1} \Pr = \C^{-1}$.  Furthermore because
only the last two equations in (\ref{struccov}) change sign, the $\P$ 
matrices satisfy $\Pr \P_T \Pr =\P_T$,  $\Pr \P_E \Pr =\P_E$,  $\Pr \P_B
\Pr =\P_B$, $\Pr \P_X \Pr =\P_X$, $\Pr \P_Y \Pr =-\P_Y$ and $\Pr \P_Z
\Pr =-\P_Z$.  In other words, the $Y$ and $Z$ spectra are odd under
parity while the $T$, $E$ and $X$ ones are even.

We now have all the necessary ingredients to show that any cross window
function that involves spectra with different parity will be
zero. These window functions will be given by the values of
$\tr[\C^{-1}\P_i \C^{-1}\P_{i'}]$. If the two $\P$-matrices
have different parities, we have 
\beq{tracepar}
\tr[\C^{-1}\Pr \P_i \Pr
\C^{-1}\Pr \P_{i'}\Pr]=-\tr[\C^{-1}\P_i \C^{-1}\P_{i'}].
\eeq 
Using the fact that $\C^{-1}$ commutes with $\Pr$ and the cyclic
property of the trace, we get 
\beq{final}
\tr[\C^{-1} \P_i
\C^{-1}\P_{i'}]=-\tr[\C^{-1}\P_i \C^{-1}\P_{i'}].
\eeq
Thus if $\P_i$ and $\P_{i'}$ have different parities, the trace  will be
zero. With our choice of $\C$, the  window functions mixing spectra with
different parities are identically  zero.
This means that there is no leakage between X and Y, between E and Z 
or between B and Z, so the only non-zero mixed
window function is $W^{\l E}_{\l' B}$.
In other words, out of the 15 potential leakages, our method eliminates all except 
that between E and B.

\bigskip
\bigskip
\bigskip

Support for this work was provided by
NASA grants NAG5-9194 and NAG5-11099,
NSF grant AST00-71213,
and two awards from the David and Lucile Packard Foundation.


\end{document}